\documentclass[aps,twocolumn,nofootinbib,showpacs, prd]{revtex4}
\usepackage{graphicx}
\usepackage{amsfonts}
\usepackage{amssymb}
\usepackage{amsbsy}
\usepackage{amsmath}
\usepackage{latexsym}
\usepackage{natbib}
\usepackage{bm}
\usepackage{subfigure}
\usepackage{color}

\def\ce{\mathrm{ce}}
\def\se{\mathrm{se}}

\def\k{\mathbf{k}}

\def\x{\mathbf{x}}
\def\y{\mathbf{y}}
\def\xiv{\vec{\xi}}

\def\k{\mathrm{k}}

\def\H{\mathcal{H}}

\def\kcut{k_{max}}
\def\lstar{l_{\star}}

\begin{document}

\title{Violation of the Kubo-Martin-Schwinger condition along a Rindler 
trajectory in polymer quantization}

\author{Golam Mortuza Hossain}
\email{ghossain@iiserkol.ac.in}

\author{Gopal Sardar}
\email{gopal1109@iiserkol.ac.in}

\affiliation{ Department of Physical Sciences, 
Indian Institute of Science Education and Research Kolkata,
Mohanpur - 741 246, WB, India }
 
\pacs{04.62.+v, 04.60.Pp}

\date{\today}

\begin{abstract}

Existence of Unruh effect is often understood from the property of two-point 
function along Rindler trajectory where it satisfies KMS condition. In 
particular, it exhibits the so-called KMS periodicity along imaginary time 
direction. Corresponding period is then identified with reciprocal of 
Unruh temperature times Boltzmann constant. We show here that the two-point 
function including leading order perturbative corrections due to polymer 
quantization, the quantization method used in loop quantum gravity, violates KMS 
condition in low-energy regime. This violation is caused by correction 
terms which are \emph{not} Lorentz invariants. Consequently, polymer corrected 
two-point function along Rindler trajectory looses its thermal interpretation. 
We discuss its implications on existence of Unruh effect in the context of 
polymer quantization.

\end{abstract}

\maketitle

\section{Introduction}

With respect to a \emph{uniformly accelerating} observer,  Fock vacuum state 
appears as a \emph{thermal} state rather than a zero-particle state.
This phenomena is referred as Unruh effect \cite{Fulling:1972md, 
Unruh:1976db,Crispino:2007eb,de2006unruh,Takagi01031986,Longhi:2011zj} and it is 
an important result of standard quantum field theory when applied to a curved 
spacetime \cite{Birrell1984quantum,wald1994quantum}. Unruh effect can be 
realized in many different ways. Firstly, one can employ the method of 
\emph{Bogoliubov transformation} in which one computes expectation value of 
number density operator associated with the accelerating observer,  in Fock 
vacuum state. The result of this computation turns out to be a blackbody 
distribution of a given temperature known as \emph{Unruh temperature}.
In this method, one needs to include contributions from 
trans-Planckian modes as seen by an \emph{inertial} observer. This particular 
aspect makes Unruh effect to be a potentially interesting arena for 
understanding and exploring implications of Planck-scale physics
\cite{Nicolini:2009dr,Padmanabhan:2009vy,Agullo:2008qb}. In 
particular, one may use it for probing a candidate theory of quantum gravity as 
trans-Planckian modes are expected to receive significant modifications from it.
While the method of Bogoliubov transformation is conceptually simpler but it 
requires sophisticated regularization techniques to deal with associated field 
theoretical divergences. Therefore, it is often imperative to understand 
the origin of Unruh effect using different methods.

In quantum statistical mechanics, it is well known that Gibbs ensemble average 
of two-point function which is in thermal equilibrium with a \emph{reservoir} 
satisfies the so-called Kubo-Martin-Schwinger (KMS) condition 
\cite{Haag:1967sg,Hugenholtz:1971yr,Bratteli:1981operator}. In particular, 
thermal two-point function exhibits \emph{periodicity with a twist} along 
imaginary time direction \cite{Fulling1987135}. Corresponding period is then 
identified with reciprocal of reservoir temperature times \emph{Boltzmann} 
constant. Subsequently, this property is used in reverse to argue that if a 
two-point function computed in a state, satisfies KMS condition then one can 
associate a thermal characteristic to the state. As an example, it can be shown 
that two-point function computed in Fock vacuum state and when viewed along the 
trajectory of a \emph{uniformly accelerating} observer, appears like a thermal 
two-point function \cite{Troost:1977dw,Troost:1978yk} as it satisfies KMS 
condition. The corresponding KMS period then leads to a reservoir temperature 
which is precisely equal to Unruh temperature.

As mentioned earlier, in the method of Bogoliubov transformation, one 
needs to include trans-Planckian modes. These modes are expected to be modified
significantly by Planck-scale physics. Therefore, one is naturally led to ask 
whether the effects from a theory of Planck scale physics can 
alter a few specific properties of two-point function even in low-energy regime. 
In particular, one can ask whether the corrections from such a theory can lead 
to a violation of KMS condition. In this paper, we perform the task in the 
context of polymer quantization.

Polymer quantization or loop quantization 
\cite{Ashtekar:2002sn,Halvorson-2004-35} is a  quantization method which is used 
in loop quantum gravity 
\cite{Ashtekar:2004eh,Rovelli2004quantum,Thiemann2007modern}. It differs from 
Schr\"odinger quantization in several aspects when applied to a 
mechanical system. In particular, apart from \emph{Planck constant} $\hbar$, it 
comes with a new dimension-full parameter. In the context of full quantum 
gravity, this new scale would correspond to \emph{Planck length}. 
Secondly, kinematical Hilbert space being non-separable, one cannot define 
both position and momentum operators simultaneously in polymer quantization but 
only one of them. Due to the non-separability of kinematical Hilbert space, the 
Stone-von Neumann uniqueness theorem is also not applicable. These aspects make 
polymer quantization unitarily \emph{inequivalent} to Schr\"odinger 
quantization \cite{Ashtekar:2002sn}. Therefore, in principle, one can get a 
different set of results from polymer quantization.

In usual derivation of two-point function in Fock space, the field 
operator is expressed in terms of \emph{creation} and \emph{annihilation} 
operators of different Fourier modes. Each of these modes behaves as a 
\emph{mechanical} system corresponding to a decoupled harmonic oscillator. 
Subsequently these modes are quantized using Schr\"odinger quantization method. In 
polymer quantization of these modes, the notions of creation and annihilation 
operators are \emph{not} readily available. Therefore, following reference
\cite{Hossain:2010eb}, we use a slightly improvised method in which 
two-point function is derived using \emph{energy spectrum} of these modes.

In section \ref{Sec:RindlerObserver}, we briefly review some aspects of a 
\emph{uniformly accelerating} observer and its associated Rindler spacetime. In 
section \ref{Sec:KMSCondition}, we revisit KMS condition in the context of 
quantum statistical mechanics. In section \ref{Sec:MasslessScalarField}, we 
consider a massless scalar field to compute two-point function in both Fock and 
polymer quantization. Then we show that polymer corrected two-point function 
along Rindler trajectory \emph{does not} satisfy KMS condition. Consequently, 
polymer corrected two-point function for Rindler observer looses its thermal 
interpretation. The results shown here is consistent with the result of 
\cite{Hossain:2014fma} where it is shown using method of Bogoliubov 
transformation that Unruh effect is absent in polymer quantization.

\section{Rindler Observer}\label{Sec:RindlerObserver}

A section of Minkowski spacetime when viewed from a \emph{uniformly 
accelerating} frame, can be described by Rindler metric. Using \emph{conformal} 
Rindler coordinates $\bar{x}^{\alpha} = (\tau,\xi,y,z) \equiv (\tau,\xiv)$ 
together with \emph{natural} units ($c=\hbar=1$), Rindler metric can be 
expressed as \cite{Rindler:1966zz}
\begin{equation}
 \label{RindlerMetric}
  ds^2 = e^{2a\xi} \left( -d\tau^2 + d\xi^2 \right) + dy^2 + dz^2
  \equiv g_{\alpha\beta}d\bar{x}^{\alpha} d\bar{x}^{\beta} ~.
\end{equation}
The parameter $a$ here denotes magnitude of \emph{acceleration} 4-vector. 
To an \emph{inertial} observer who uses the coordinates $x^{\mu} = 
(t,x,y,z) \equiv (t,\x)$, the Minkowski metric would appear as $ds^2 =
\eta_{\mu\nu}dx^{\mu}dx^{\nu} = - dt^2 + dx^2 + dy^2 + dz^2$.
If the uniformly accelerating observer, referred as Rindler observer, moves 
along $+ve$ x-axis with respect to the inertial observer then their coordinates 
are related to each other as
\begin{equation}
 \label{RindlerMinkowskiRelation}
 t =  \frac{1}{a} e^{a\xi} \sinh a\tau ~,~~~
 x =  \frac{1}{a} e^{a\xi} \cosh a\tau ~.
\end{equation}
The $y$ and $z$ coordinates are related trivially. 
The equation (\ref{RindlerMinkowskiRelation}) makes it clear that
Rindler spacetime covers only a wedge-shaped region, referred as
\emph{Rindler wedge}, of Minkowski spacetime.

\section{KMS condition}\label{Sec:KMSCondition}

In quantum statistical mechanics, Gibbs ensemble average of an observable 
$\hat{O}$ which is in thermal equilibrium with a \emph{reservoir} of 
temperature $T$, can be expressed as
\begin{equation}
 \label{EnsembleAverageDef}
\langle \hat{O} \rangle_{\beta} =  Z^{-1} 
Tr\left[ e^{-\beta \hat{H}} \hat{O}\right] ~,
\end{equation}
where $\beta = 1 /k_B T$, $\hat{H}$ is the associated Hamiltonian operator 
and $Z = Tr\left[ e^{-\beta \hat{H}}\right]$ is the corresponding 
\emph{partition function}. Here $k_B$ refers to \emph{Boltzmann constant}. If we 
consider the observable to be $\hat{O} = \hat{\phi}(\tau,\xiv)~ 
\hat{\phi}(\tau',\xiv')$ where $\hat{\phi}(\tau,\xiv)$ is a given field 
operator, then thermal two-point function associated with the field can 
be expressed as
\begin{equation}
 \label{EnsembleAverageTwoPoint1}
\langle \hat{\phi}(\tau,\xiv) \hat{\phi}(\tau',\xiv') \rangle_{\beta} = 
 Z^{-1} 
Tr\left[ e^{-\beta \hat{H}} \hat{\phi}(\tau,\xiv) 
\hat{\phi}(\tau',\xiv')
\right] ~.
\end{equation}
Using time evolution of the field operator as 
$\hat{\phi}(\tau,\xiv) = e^{i\hat{H}\tau} \hat{\phi}(0,\xiv) e^{-i\hat{H}\tau}$ 
and invariance of trace under \emph{cyclic permutation}, one can easily show 
that
\begin{equation}
 \label{KMSConditionGeneral}
\langle \hat{\phi}(\tau,\xiv) \hat{\phi}(\tau',\xiv') \rangle_{\beta} = 
\langle \hat{\phi}(\tau',\xiv') \hat{\phi}(\tau+i\beta,\xiv)  
\rangle_{\beta} ~.
\end{equation}
The equation (\ref{KMSConditionGeneral}) is commonly referred as 
Kubo-Martin-Schwinger (KMS) condition 
\cite{Hugenholtz:1971yr,Bratteli:1981operator}. One may note that relative 
position of field operators in two sides are interchanged. Therefore, 
corresponding periodicity of two-point function along \emph{imaginary} time 
direction (\ref{KMSConditionGeneral}) is often referred as \emph{periodicity 
with a twist} \cite{Fulling1987135}.

\section{Massless scalar field}
\label{Sec:MasslessScalarField}

The dynamics of a massless scalar field $\Phi(x)$ in Minkowski spacetime is 
governed by the action
\begin{equation}
 \label{ScalarActionMinkowski}
 S_{\Phi} = \int d^4x \left[ - \frac{1}{2} \sqrt{-\eta} \eta^{\mu\nu}  
 \partial_{\mu} \Phi(x)  \partial_{\nu} \Phi(x) \right] ~.
\end{equation}
In this paper, we are mainly interested in employing polymer quantization scheme
which is a canonical quantization method. Therefore, it is necessary for us to 
compute the Hamiltonian associated with the action 
(\ref{ScalarActionMinkowski}). By considering \emph{spatial} hyper-surfaces 
which are labeled by $t$, one can compute corresponding Hamiltonian as
\begin{equation}\label{SFHamGen}
H_{\Phi}  =  \int d^3\x \left[ \frac{\Pi^2}{2\sqrt{q}} +
\frac{\sqrt{q}}{2} q^{ab} \partial_a\Phi \partial_b\Phi
\right] ~,
\end{equation}
where $q_{ab}$ is the metric on spatial hyper-surfaces. Poisson bracket 
between the field $\Phi = \Phi(t,\x)$ and the conjugate field momentum $\Pi = 
\Pi(t,\x)$ are given by
\begin{equation}\label{PositionSpacePB}
\{\Phi(t,\x), \Pi(t,\y)\} = \delta^{3}(\x-\y) ~.
\end{equation}

\subsection{Fourier modes}

We define Fourier modes for the scalar field and its conjugate momentum 
in Minkowski spacetime as
\begin{equation}\label{FourierModesDef}
\Phi = \frac{1}{\sqrt{V}} \sum_{\k} \tilde{\phi}_{\k}(t) e^{i \k\cdot\x} ,~
\Pi  = \frac{1}{\sqrt{V}} \sum_{\k} \sqrt{q} ~\tilde{\pi}_{\k}(t) 
e^{i \k\cdot\x},
\end{equation}
where $V=\int d^3\x \sqrt{q}$ is the spatial volume. For Minkowski spacetime, 
the spatial volume diverges given the space is non-compact. Therefore, in order 
to avoid dealing with divergent quantity, one considers a fiducial box of 
finite volume. In such situation, Kronecker delta and Dirac delta are expressed 
as
$\int d^3\x \sqrt{q} ~e^{i (\k-\k')\cdot \x} = V \delta_{\k,\k'}$ and 
$\sum_{\k} e^{i \k\cdot (\x-\y)} = V \delta^3 (\x-\y)/\sqrt{q}$. 
In terms of Fourier modes (\ref{FourierModesDef}), the Hamiltonian
(\ref{SFHamGen}) can be expressed as $H_{\Phi} = \sum_{\k} \H_{\k}$, where
Hamiltonian density for the $\k-$th mode is
\begin{equation}\label{SFHamFourierMinkowski}
\H_{\k} = \frac{1}{2} \tilde{\pi}_{-\k} \tilde{\pi}_{\k} +
\frac{1}{2} |\k|^2 \tilde{\phi}_{-\k}\tilde{\phi}_{\k}  ~.
\end{equation}
Corresponding Poisson bracket is given by
\begin{equation}\label{Minkowski:MomentumSpacePB}
\{\tilde{\phi}_{\k}, \tilde{\pi}_{-\k'}\} =  \delta_{\k,\k'}
~.
\end{equation}
By redefining \emph{complex-valued} modes $\tilde{\phi}_{\k}$ and momenta 
$\tilde{\pi}_{\k}$ in terms of real-valued functions $\phi_{\k}$ and
$\pi_{\k}$ such that \emph{reality condition} of the scalar field $\Phi$ is
ensured, one can express Hamiltonian density and Poisson bracket as
\begin{equation}\label{SFHamFourierMinkowskiReal}
\H_{\k} = \frac{1}{2} \pi_{\k}^2 + \frac{1}{2} |\k|^2 \phi_{\k}^2
~~~;~~~ \{\phi_{\k},\pi_{\k'}\} =  \delta_{\k,\k'} ~.
\end{equation}
This is the usual Hamiltonian for a set of decoupled harmonic oscillators. 
If we denote energy spectrum of these quantum oscillators as 
$\hat{\H}_{\k}|n_{\k}\rangle = E_n^{(\k)}|n_{\k}\rangle$ then 
corresponding vacuum state can be expressed as $|0\rangle=\Pi_{\k}\otimes 
|0_\k\rangle$.

\subsection{Two-point function in Minkowski spacetime}

In position space two-point function using Minkowski coordinates
can be written as
\begin{equation}
\label{MinkowskiTwoPointDef}
G(x,x') \equiv \langle 0|\hat{\Phi}(x) \hat{\Phi}(x')|0\rangle
= \langle 0|\hat{\Phi}(t,\x) \hat{\Phi}(t',\x')|0\rangle
\end{equation}
where $|0\rangle$ denotes corresponding vacuum state. Using definition of 
Fourier modes, one can express two-point function (\ref{MinkowskiTwoPointDef}) 
as
\begin{equation}
\label{MinkowskiTwoPointDef2}
G(x,x') = \frac{1}{V} \sum_{\k} D_{\k}(t,t') ~e^{i {\k}\cdot(\x-\x')} ,
\end{equation}
where the matrix element can be written as
\begin{equation}\label{DkDefinition}
D_{\k}(t,t') = 
\langle 0_{\k}| e^{i\hat{\H}_{\k}t} \hat{\phi}_{\k} e^{-i\hat{\H}_{\k}t}
e^{i\hat{\H}_{\k}t'} \hat{\phi}_{\k} e^{-i\hat{\H}_{\k}t'}
|0_{\k}\rangle.
\end{equation}
We now remove the fiducial volume by taking the limit $V \to \infty$. This 
essentially requires that we replace the sum $\frac{1}{V} \sum_{\k}$ by an 
integration $\int \frac{d^3\k}{(2\pi)^3}$. Thanks to the definition of Fourier 
modes (\ref{FourierModesDef}), the equation (\ref{SFHamFourierMinkowskiReal}) is 
independent of the fiducial volume. Therefore, two-point function 
(\ref{MinkowskiTwoPointDef2}) becomes
\begin{equation}
\label{MinkowskiTwoPointIntegralDef}
G(x,x') = \int \frac{d^3\k}{(2\pi)^3} ~  D_{\k}(t,t') 
~e^{i {\k}\cdot(\x-\x')} ~.
\end{equation}
Using energy spectrum and expanding the state $\hat{\phi}_{\k} |0_{\k}\rangle$ 
in the basis of energy eigenstates as 
$\hat{\phi}_{\k}|0_{\k}\rangle = \sum_{n} c_n |n_{\k}\rangle$, one can
reduce the matrix element as
\begin{equation}
\label{DkFunctionGeneral}
D_{\k}(t-t') \equiv D_{\k}(t,t') = \sum_{n} |c_n|^2 e^{-i\Delta E_n (t-t')},
\end{equation}
where $\Delta E_n \equiv E_n^{(\k)} - E_0^{(\k)}$ and
$c_n = \langle n_{\k}| \hat{\phi}_{\k} |0_{\k}\rangle$.
We shall see later that $\k$ dependence of $D_{\k}(t-t')$ is through $|\k|$. So
above integration can be evaluated conveniently by using \emph{polar 
coordinates} in momentum space as
\begin{equation}
\label{KGPropagator}
G(x,x') =  \int \frac{k^2 dk}{4\pi^2} D_{k}(\Delta t) 
\int \sin\theta d\theta ~e^{i k |\Delta \x| \cos\theta }  ,
\end{equation}
where $k = |\k|$, $\Delta \x = \x-\x'$ and $\Delta t = t-t'$. 
By carrying out $\theta$ integration one can express the two-point function as
\begin{equation}
\label{KGPropagatorDiffPM}
G(x,x') = G_{+} - G_{-} ~,
\end{equation}
where 
\begin{equation}
\label{GPMDefinition}
G_{\pm} =  \frac{i}{4\pi^2|\Delta\x|} \int dk ~k D_{k}(\Delta t) 
~e^{\mp i k |\Delta \x|} ~.
\end{equation}

\subsubsection{Fock quantization}

In Fock space, Fourier modes, as represented by equation 
(\ref{SFHamFourierMinkowskiReal}), are quantized using Schr\"odinger 
quantization method. It is straightforward to compute corresponding energy 
spectrum and the coefficients $c_n$ as
\begin{equation}\label{FockSpectrum}
E^n_{\k} = \left(n+\frac{1}{2}\right)|\k| ~~;~~ 
\Delta E_n = n |\k|   ~~;~~
 c_n = \frac{\delta_{1,n}}{\sqrt{2|\k|}} ~.
\end{equation}
The expressions as given in (\ref{FockSpectrum}), led to a simpler form of the 
matrix element 
\begin{equation}\label{KGPropagator3}
D_{\k}(\Delta t) =  \frac{1}{2|\k|} e^{ -i |\k| \Delta t} ~.
\end{equation}
Along a timelike trajectory $(\Delta t \pm |\Delta \x|)$ is always 
\emph{positive} when $\Delta t>0$. So by defining a new variable $u = k (\Delta 
t \pm |\Delta \x|)$, we can transform the expression of $G_{\pm}$ 
(\ref{GPMDefinition}) in Fock space as
\begin{equation}
\label{GPM}
G_{\pm} = \frac{i}{8\pi^2|\Delta\x|} ~\frac{I_0}{(\Delta t \pm |\Delta \x|)} ~,
\end{equation}
where $I_n = \int_0^{\infty} du~ u^n e^{-iu}$. We need to evaluate $I_n$  only 
for $n=0$ here. However for later convenience, we evaluate $I_n$ for $n 
\ge 0$. Firstly we note that integral expression of $I_n$ is formally 
\emph{divergent}. Therefore, in order to regulate it, we introduce a 
\emph{small, positive} parameter $\epsilon$ in the integrand such that 
regulated expression of $I_n$ becomes
\begin{equation}
 \label{InFunction}
I_n^{\epsilon} = \int_0^{\infty} du~ u^n e^{-iu - \epsilon u} = 
\frac{\Gamma(n+1)}{(i+\epsilon)^{n+1}} ~~.
\end{equation}
Using regulated $I_n^{\epsilon}$, it is straightforward to simplify the 
two-point function as
\begin{equation}
\label{FockTwoPointFunction}
G(x,x') = \frac{(1 - i\epsilon)^{-1}}{4\pi^2 \Delta x^2} ~,
\end{equation}
where $\Delta x^2 = - \Delta t^2 + |\Delta \x|^2$ is Lorentz invariant 
spacetime interval.

\subsubsection{Polymer quantization}

Polymer quantization comes with a new dimension-full parameter say $\lstar$ 
which is analogous of Planck length in full quantum gravity. For each Fourier 
mode, one may define a \emph{dimensionless} parameter $g = |\k|~\lstar$. It 
signifies the scale of polymer corrections. Energy eigenvalues for $\k-$th 
oscillator in polymer quantization \cite{Hossain:2010eb} are given by 
\begin{equation}
 \label{EigenValueMCFRelation}
 \frac{E_{\k}^{2n}}{|\k|} = \frac{1}{4g} + \frac{g}{2} ~A_n(g)  ~,~
 \frac{E_{\k}^{2n+1}}{|\k|} = \frac{1}{4g} + \frac{g}{2} B_{n+1}(g) ~,
\end{equation}
where $n\ge0$, $A_n$ and $B_n$ are \emph{Mathieu characteristic value
functions}. The corresponding energy eigenfunctions are 
$\psi_{2n}(v) = \ce_n(1/4g^2,v)/\sqrt{\pi}$ and 
$\psi_{2n+1}(v) = \se_{n+1}(1/4g^2,v)/\sqrt{\pi}$ where 
$v = \pi_{\k} \sqrt{\lstar} + \pi/2 $.
The functions $\ce_n$ and $\se_n$ are solutions to \emph{Mathieu
equation}. They are referred as elliptic cosine and sine functions
respectively \cite{Abramowitz1964handbook}.
We may emphasize here that in order to arrive at these $\pi$-periodic and 
$\pi$-antiperiodic states in $v$, one invokes superselection rules motivated 
by the symmetry of associated potential in energy eigenvalue equation. In 
particular, the potential when expressed in terms of the variable 
$v$, is periodic with period $\pi$. The corresponding energy eigenvalues
are then expressed exactly in terms of Mathieu characteristic value
functions $A_n$ and $B_n$. Having exact energy spectrum allows one
to study different aspects of the system analytically. Furthermore, without 
imposition of superselection rules some aspects of the given system are 
known to be problematic \cite{Barbero:2013lia}. In particular, it is shown that 
even basic notions of statistical entropy, canonical partition function for 
such a system become ill-defined due to the effective infinite degeneracy of 
the energy eigenvalues in polymer quantization, unless one invokes some 
superselection rules \cite{Barbero:2013lia}.

Unlike Fock space, matrix element $D_{\k}(\Delta t)$ 
(\ref{DkFunctionGeneral}) does not appear to be exactly summable given 
all $c_{4n+3} = i \sqrt{\lstar} \int_0^{2\pi} 
\psi_{4n+3} \partial_v\psi_{0} dv$ (for $n=0, 1, 2, \ldots$) are 
non-vanishing in polymer 
quantization.  However, asymptotic properties of Mathieu functions are 
well known and they may be used to evaluate the matrix element approximately. 
Following reference \cite{Hossain:2010eb}, we note that for low-energy 
modes (compared to $\lstar$, \emph{i.e.} 
$g\ll 1$)
\begin{equation}
\frac{\Delta E_{4n+3}}{|\k|} = (2n+1) - \frac{(4n+3)^2-1}{16} g  + 
\mathcal{O} \left( g^2 \right) ~,
\end{equation}
for $n \ge 0$, and 
\begin{equation}
\label{SmallgDeltaE4n+3}
  c_3 = \frac{i}{\sqrt{2|\k|}} \left[1 
  +\mathcal{O}\left(g\right) \right] ~,~
  \frac{c_{4n+3}}{c_{3}} = \mathcal{O}\left(g^n\right),
\end{equation}
for $n >0$. On the other hand for trans-Planckian modes (\emph{i.e.} 
$g\gg1$), we note that
\begin{equation}
 \label{LargegDeltaE4n+3}
 \frac{\Delta E_{4n+3} }{|\k|}  =  2(n+1)^2 g + 
  \mathcal{O}\left(\frac{1}{g^3}\right) ,
\end{equation}
  for $n \ge 0$, and 
\begin{equation}
  \label{LargegC4n+3}
  c_{3} =  i\sqrt{\frac{g}{2|\k|}} \left[\frac{1}{4g^2} +
  \mathcal{O}\left(\frac{1}{g^6}\right) \right],
  \frac{c_{4n+3}}{c_{3}} = \mathcal{O}\left(\frac{1}{g^{2n}}\right),
\end{equation}
for $n > 0$.  
So in both asymptotic cases, leading contributions to the expression 
of $D_{\k}(\Delta t)$ (\ref{DkFunctionGeneral}) comes from $c_{3}$ term
compared to other non-vanishing $c_{4n+3}$ terms. Therefore, one can 
approximate matrix element $D_{\k}(\Delta t)$ in polymer quantization as
\begin{equation}
\label{DkPolyApprox1}
D_{\k}^{poly}(\Delta t) \simeq  |c_3|^2 e^{ -i \Delta E_3 \Delta t} ~.
\end{equation}
Further, we note that for low-energy modes ($g \ll 1$), modulus of the 
\emph{integrand} in equation (\ref{GPMDefinition}) varies as
\begin{equation}
\label{DkAmpltitudeInfrared}
|k~D_{\k}^{poly}(\Delta t)|  = \frac{1}{2} \left[1 + 
\mathcal{O}\left( g \right)  \right] ~,
\end{equation}
whereas for trans-Planckian modes ($g \gg 1$), modulus of the 
integrand varies as
\begin{equation}
\label{DkAmpltitudeUltraviolet}
|k~D_{\k}(\Delta t)| = \frac{1}{32 g^3} \left[ 1 + 
\mathcal{O}\left(\frac{1}{g^4}\right) \right] \ll 1 ~.
\end{equation}
So to evaluate the expression of $G_{\pm}$ (\ref{GPMDefinition}) in 
polymer quantization, one may restrict the integration along $k$ up to a 
pivotal $\kcut$. In other words, polymer corrected super-Planckian modes 
\emph{do not contribute} significantly towards two-point function as expected. 
Nevertheless, each low-energy mode that does contribute to the two-point 
function, comes with \emph{perturbative} polymer corrections. 

In order to understand the implications of each correction terms separately, we 
parameterize the perturbative polymer corrections to energy spectrum and 
coefficient $c_3$ as
\begin{equation}
\Delta E_{3}^{\star} = k\left(1 + \delta_{E3}~ \lstar k\right) ~;~
|c_3|^2_{\star} = \frac{1}{2k} \left(1 + 2\delta_{c3} ~\lstar k\right) ~,
\end{equation}
where $\delta_{E3}$ and $\delta_{c3}$ are numerical constants signifying 
the deviation from Fock quantization and they carry the label of their 
origin. We now express $G_{\pm}$ (\ref{GPMDefinition}) with perturbative 
polymer corrections as
\begin{equation}
\label{GPMPolyDef}
G_{\pm}^{poly} \simeq \frac{i}{4\pi^2|\Delta\x|}  
\int_0^{\kcut} dk ~k ~D_{k}^{\star}(\Delta t) ~e^{ \mp  i k|\Delta \x|} ~,
\end{equation}
where $D_{\k}^{\star}(\Delta t) = |c_3|^2_{\star} e^{ -i \Delta E_3^{\star} 
\Delta t}$.
By defining a new variable $u = k(\Delta t \pm |\Delta \x| + \delta_{E3} \lstar 
k\Delta t)$, one can transform the integral (\ref{GPMPolyDef}) as
\begin{eqnarray}
\label{GPMPolymer}
 && G_{\pm}^{poly}(x,x') \simeq \frac{i}{8\pi^2|\Delta\x|} 
\left[ \frac{I_0(u_0^{\pm})}{(\Delta t \pm |\Delta \x|)} 
~+   \right. \nonumber \\
&& \left. 2 \lstar~ I_1(u_0^{\pm}) \left\{
\frac{\delta_{c3} }{(\Delta t \pm |\Delta \x|)^2} 
- \frac{\delta_{E3} ~\Delta t}{(\Delta t \pm |\Delta \x|)^3}
\right\}
\right] ,~~~
\end{eqnarray}
where $I_n(u_0^{\pm}) = \int_0^{u_0^{\pm}} du~ u^n e^{-iu}$
with $u_0^{\pm} = u_{|\kcut}$.
In order to perform perturbative expansion in (\ref{GPMPolymer}),
we have assumed $\lstar \kcut\Delta t \ll (\Delta t \pm |\Delta \x|)$.
For a given characteristic length scale $(\Delta t \pm |\Delta \x|)$ 
which is associated with two-point function, we can choose appropriately 
large $\kcut$, such that $u_0^{\pm} \gg 1$ yet $\kcut \lstar \ll 1$. 
This in turn restricts the domain of perturbation in which 
$\lstar \Delta t \ll (\Delta t \pm |\Delta \x|)^2$. By introducing a small 
parameter $\epsilon$ again as a regulator, we can evaluate the integral 
approximately as
\begin{equation}
 \label{InFunctionPoly}
 I_n^{\epsilon}(u_0^{\pm}) \approx  \frac{\Gamma(n+1)}{(i + \epsilon)^{n+1}} 
~,
\end{equation}
where we have ignored terms which are exponentially small as $u_0^{\pm}$ 
is \emph{large but finite}. It is now straightforward to express polymer 
corrected two-point function, including leading order \emph{perturbative 
correction terms} as
\begin{equation}
\label{PolymerCorrectedTwoPointFunction}
G^{poly}(x,x') \simeq \frac{(1-i\epsilon)^{-1}}{4\pi^2 \Delta x^2}
\left[1 + \frac{2 i ~ \delta^{poly} ~ \lstar \Delta t }
{(1-i\epsilon) \Delta x^2}
\right] ~,
\end{equation}
where $\delta^{poly} = 2 ~\delta_{c3} + \delta_{E3}
\left[1 + 4 (\Delta t^2/\Delta x^2) \right]$.
We note that in the limit $\lstar \to 0$, polymer corrected 
two-point function (\ref{PolymerCorrectedTwoPointFunction}) reduces to Fock 
space two-point function (\ref{FockTwoPointFunction}). Leading order polymer 
correction to the two-point function is of $\mathcal{O} \left(\lstar\right)$. 
However, being purely \emph{imaginary} number, leading order polymer 
corrections to the modulus of two-point function which may have direct 
observational relevance, are of $\mathcal{O} \left(\lstar^2\right)$. We also 
note that leading order polymer correction terms explicitly \emph{violate} 
Lorentz invariance.

\section{Rindler Two-point function}\label{RindlerTwoPoint}

Let us consider a thermal detector which is moving along Rindler observer and 
is located at a spatial position $\xiv_0$ in Rindler frame. If one assumes that 
the detector is at thermal equilibrium with a reservoir of temperature T then 
thermal two-point function (\ref{EnsembleAverageTwoPoint1}) associated with the 
detector is 
\begin{equation}
 \label{ThermalTwoPoint}
\mathcal{G}(\tau,\tau') \equiv
\langle \hat{\phi}(\tau,\xiv_0) \hat{\phi}(\tau',\xiv_0) \rangle_{\beta} ~.
\end{equation}
For such a detector, KMS condition (\ref{KMSConditionGeneral}) then becomes
\begin{equation}
\label{KMSConditionRindler1}
\mathcal{G}(\tau,\tau') = \mathcal{G}(\tau', \tau + i\beta) ~.
\end{equation}
For simplicity we assume that the detector is located at the origin of Rindler 
frame \emph{i.e.} $\xiv_0 = (0,0,0)$ and we choose $\tau'=0$. Using time 
evolution of the field operator $\hat{\phi}(\tau,\xiv_0)$, one can show that 
$\mathcal{G}(\tau,0) = \mathcal{G}(0,-\tau)$. Therefore by defining 
$\mathcal{G}(\tau) \equiv \mathcal{G}(0,\tau)$, we can further simplify KMS 
condition (\ref{KMSConditionRindler1}) as
\begin{equation}
\label{KMSConditionRindler2}
\mathcal{G}(-\tau) = \mathcal{G}(\tau + i\beta) ~.
\end{equation}

Now using equation (\ref{RindlerMinkowskiRelation}), trajectory of the detector 
can be expressed as $x_d(\tau) = (\sinh a\tau/a, \cosh a\tau/a, 0, 0)$. So
Fock space two-point function (\ref{FockTwoPointFunction}) along the trajectory 
of the detector can be written as
\begin{equation}
\label{FockTwoPointFunctionInRindler}
G(\tau) \equiv G\left(x_d(\tau),x_d(0)\right) = 
\frac{a^2(1-i\epsilon)^{-1}}{8\pi^2(1-\cosh a\tau)} ~.
\end{equation}
Clearly, two-point function (\ref{FockTwoPointFunctionInRindler}) 
satisfies the condition $G(-\tau) = G(\tau + i \beta)$ with $\beta = 2\pi/a$. 
In other words, two-point function computed in Fock vacuum but when viewed 
from a uniformly accelerating frame \emph{i.e.} Rindler frame, appears to 
satisfy KMS condition (\ref{KMSConditionRindler2}). Therefore, one may conclude 
that with respect to Rindler observer, Fock vacuum appears like a 
\emph{thermal reservoir} of temperature $T=a/2\pi k_B$ which is
precisely equal to Unruh temperature.

Following similar setup as in Fock space, polymer corrected two-point 
function (\ref{PolymerCorrectedTwoPointFunction}) along the trajectory 
of the detector in Rindler frame can be expressed, within the domain
of perturbation given by $\lstar \ll \Delta t(\tau)$ and 
$a^2 \lstar \Delta t(\tau) \ll 1$, as
\begin{equation}
\label{PolymerTwoPointFunctionInRindler}
G^{poly}(\tau) = \frac{a^2 (1-i\epsilon)^{-1}}
{8\pi^2 (1-\cosh a\tau)}
\left[1 + \Delta G^{(1)}_{\star} + \mathcal{O}(\lstar^2)\right] ,
\end{equation}
where
\begin{equation}
\label{PolymerTwoPointFunctionInRindlerDelta}
\Delta G^{(1)}_{\star} = \frac{i \lstar a \sinh a\tau
[2 \delta_{c3} - \delta_{E3}(1 + 2\cosh a\tau)]
}
{(1-i\epsilon)(1-\cosh a\tau)} ~.
\end{equation}
We note that unlike in the case of Fock space, polymer corrected 
two-point function (\ref{PolymerTwoPointFunctionInRindler}) does not satisfy 
KMS condition (\ref{KMSConditionRindler2}) given
\begin{equation}
\label{KMSViolationPoly}
 G^{poly}(-\tau) \ne G^{poly}(\tau + i \beta) ~.
\end{equation}
Therefore, polymer corrected two-point function along Rindler trajectory looses 
its thermal interpretation. From equation 
(\ref{PolymerTwoPointFunctionInRindler}, 
\ref{PolymerTwoPointFunctionInRindlerDelta}), it is clear that this violation of 
KMS condition occurs precisely due to the polymer correction terms. In 
particular, polymer correction term involving $\delta_{E3}$ makes dominant 
contribution to the violation of KMS condition as it contains $\cosh a\tau$ 
term. Contribution from the term involving $\delta_{c3}$ is comparatively 
negligible. In other words, polymer corrections to \emph{energy spectrum} of 
Fourier modes is the primary cause that leads to the violation of KMS condition. 
This violation also does not depend on any specific numerical values of 
$\delta_{c3}$ and $\delta_{E3}$. KMS condition is restored if one takes the 
limit $\lstar \to 0$.

We may note here that unlike Fock quantization there are infinitely 
many non-vanishing coefficients $c_n$ in polymer quantization. Further, the 
energy gaps $\Delta E_n$ depend \emph{non-linearly} on $n$. Therefore, 
although we have used a particular superselected sector, given the 
form of equation (\ref{DkFunctionGeneral}), it is highly unlikely that any 
other choice of superselection could have preserved KMS condition.

\section{Discussions}

In quantum statistical mechanics, Gibbs ensemble average of a two-point 
function which is in equilibrium with a thermal reservoir, satisfies KMS 
condition. This property is often used in reverse to argue that if a two-point 
function computed in a given state satisfies KMS condition then the state 
appears as a thermal state to the concerned observer. As an example, 
expectation value of number density operator in Fock vacuum state with respect 
to a \emph{uniformly accelerating} observer \emph{i.e.} Rindler observer, turns 
out to be a Plank distribution with temperature $T=a/2\pi k_B$. One can arrive 
at the same conclusion from the property of two-point function associated with 
Rindler observer. In particular, two-point function along Rindler trajectory 
satisfies KMS condition with KMS period along imaginary time being $\beta = 
2\pi/a$. This in turns implies corresponding reservoir temperature to be 
$T=a/2\pi k_B$.

In this paper, by computing leading order perturbative corrections from
polymer quantization, we have shown that polymer corrected two-point function 
along Rindler trajectory violates KMS condition. Consequently, corresponding 
two-point function looses its thermal interpretation. This violation is caused 
by polymer correction terms which are \emph{not} Lorentz invariants. 
Intuitively, this loss of thermal characteristic can also be understood
from the following arguments. A non-periodic function can be viewed as a 
periodic function but with a infinite period. In this sense, the `loss of KMS 
periodicity with a twist' can be viewed as if KMS period $\beta$ is becoming 
infinity. This in turn implies that corresponding temperature 
$T (\sim 1/\beta)$ is becoming zero. Therefore, the result shown here is 
consistent with the result of \cite{Hossain:2014fma} where it is shown using 
method of Bogoliubov transformation that Unruh effect is absent in polymer 
quantization. In both methods, polymer correction to the \emph{energy 
spectrum} of Fourier modes that leads to the loss of thermal characteristic.

In reference \cite{Rovelli:2014gva}, the author has claimed that since LQG 
would reproduce Fock space two-point function at \emph{zeroth order} then it 
should satisfy KMS condition. Therefore, `LQG predicts Unruh 
effect' \cite{Rovelli:2014gva}. This argument is \emph{flawed} as it tries to 
completely ignore all possible polymer corrections involving the new scale 
$\lstar$. In fact, we have shown here that violation of KMS condition occurs 
precisely due to polymer correction terms. Reproduction of Fock-space two-point 
function at zeroth order is a minimum requirement of viability and it does 
not imply that polymer quantization would predict exactly same physics as in 
Fock space.

We emphasize that the violation of KMS condition in Rindler frame as shown 
here, may not necessarily imply that there will be a violation of KMS condition 
in the context of black hole spacetime. There are several properties and steps 
which have been used here may not go through in case of a black hole spacetime 
\cite{Campo:2010fz}. 
Nevertheless, it would be interesting to carry out a similar analysis for a 
black hole space time. Based on the results shown here, we would like to 
reiterate that experimental detection of Unruh effect potentially can be used to 
either verify or rule out certain candidate theory of quantum gravity which 
affects matter quantization as described here. We conclude by acknowledging that 
several experimental proposals have been made in literature to test Unruh effect 
in laboratory \cite{Schutzhold:2006gj,Schutzhold:2008zza,Aspachs:2010hh}.

\begin{acknowledgments}
We would like to thank Ritesh Singh for discussions. GS would like to thank UGC 
for supporting this work through a doctoral fellowship. 
\end{acknowledgments}


\begin{thebibliography}{32}
\expandafter\ifx\csname natexlab\endcsname\relax\def\natexlab#1{#1}\fi
\expandafter\ifx\csname bibnamefont\endcsname\relax
  \def\bibnamefont#1{#1}\fi
\expandafter\ifx\csname bibfnamefont\endcsname\relax
  \def\bibfnamefont#1{#1}\fi
\expandafter\ifx\csname citenamefont\endcsname\relax
  \def\citenamefont#1{#1}\fi
\expandafter\ifx\csname url\endcsname\relax
  \def\url#1{\texttt{#1}}\fi
\expandafter\ifx\csname urlprefix\endcsname\relax\def\urlprefix{URL }\fi
\providecommand{\bibinfo}[2]{#2}
\providecommand{\eprint}[2][]{\url{#2}}

\bibitem[{\citenamefont{Fulling}(1973)}]{Fulling:1972md}
\bibinfo{author}{\bibfnamefont{S.~A.} \bibnamefont{Fulling}},
  \bibinfo{journal}{Phys.Rev.} \textbf{\bibinfo{volume}{D7}},
  \bibinfo{pages}{2850} (\bibinfo{year}{1973}).

\bibitem[{\citenamefont{Unruh}(1976)}]{Unruh:1976db}
\bibinfo{author}{\bibfnamefont{W.}~\bibnamefont{Unruh}},
  \bibinfo{journal}{Phys.Rev.} \textbf{\bibinfo{volume}{D14}},
  \bibinfo{pages}{870} (\bibinfo{year}{1976}).

\bibitem[{\citenamefont{Crispino et~al.}(2008)\citenamefont{Crispino, Higuchi,
  and Matsas}}]{Crispino:2007eb}
\bibinfo{author}{\bibfnamefont{L.~C.} \bibnamefont{Crispino}},
  \bibinfo{author}{\bibfnamefont{A.}~\bibnamefont{Higuchi}}, \bibnamefont{and}
  \bibinfo{author}{\bibfnamefont{G.~E.} \bibnamefont{Matsas}},
  \bibinfo{journal}{Rev.Mod.Phys.} \textbf{\bibinfo{volume}{80}},
  \bibinfo{pages}{787} (\bibinfo{year}{2008}), \eprint{arXiv:0710.5373}.

\bibitem[{\citenamefont{De~Bi{\`e}vre and Merkli}(2006)}]{de2006unruh}
\bibinfo{author}{\bibfnamefont{S.}~\bibnamefont{De~Bi{\`e}vre}}
  \bibnamefont{and} \bibinfo{author}{\bibfnamefont{M.}~\bibnamefont{Merkli}},
  \bibinfo{journal}{Classical and Quantum Gravity}
  \textbf{\bibinfo{volume}{23}}, \bibinfo{pages}{6525} (\bibinfo{year}{2006}).

\bibitem[{\citenamefont{Takagi}(1986)}]{Takagi01031986}
\bibinfo{author}{\bibfnamefont{S.}~\bibnamefont{Takagi}},
  \bibinfo{journal}{Progress of Theoretical Physics Supplement}
  \textbf{\bibinfo{volume}{88}}, \bibinfo{pages}{1} (\bibinfo{year}{1986}).

\bibitem[{\citenamefont{Longhi and Soldati}(2011)}]{Longhi:2011zj}
\bibinfo{author}{\bibfnamefont{P.}~\bibnamefont{Longhi}} \bibnamefont{and}
  \bibinfo{author}{\bibfnamefont{R.}~\bibnamefont{Soldati}},
  \bibinfo{journal}{Phys.Rev.} \textbf{\bibinfo{volume}{D83}},
  \bibinfo{pages}{107701} (\bibinfo{year}{2011}), \eprint{arXiv:1101.5976}.

\bibitem[{\citenamefont{Birrell and Davies}(1984)}]{Birrell1984quantum}
\bibinfo{author}{\bibfnamefont{N.~D.} \bibnamefont{Birrell}} \bibnamefont{and}
  \bibinfo{author}{\bibfnamefont{P.~C.~W.} \bibnamefont{Davies}},
  \emph{\bibinfo{title}{Quantum fields in curved space}}, \bibinfo{number}{7}
  (\bibinfo{publisher}{Cambridge university press}, \bibinfo{year}{1984}).

\bibitem[{\citenamefont{Wald}(1994)}]{wald1994quantum}
\bibinfo{author}{\bibfnamefont{R.~M.} \bibnamefont{Wald}},
  \emph{\bibinfo{title}{Quantum field theory in curved spacetime and black hole
  thermodynamics}} (\bibinfo{publisher}{University of Chicago Press},
  \bibinfo{year}{1994}).

\bibitem[{\citenamefont{Nicolini and Rinaldi}(2011)}]{Nicolini:2009dr}
\bibinfo{author}{\bibfnamefont{P.}~\bibnamefont{Nicolini}} \bibnamefont{and}
  \bibinfo{author}{\bibfnamefont{M.}~\bibnamefont{Rinaldi}},
  \bibinfo{journal}{Phys.Lett.} \textbf{\bibinfo{volume}{B695}},
  \bibinfo{pages}{303} (\bibinfo{year}{2011}), \eprint{arXiv:0910.2860}.

\bibitem[{\citenamefont{Padmanabhan}(2010)}]{Padmanabhan:2009vy}
\bibinfo{author}{\bibfnamefont{T.}~\bibnamefont{Padmanabhan}},
  \bibinfo{journal}{Rept.Prog.Phys.} \textbf{\bibinfo{volume}{73}},
  \bibinfo{pages}{046901} (\bibinfo{year}{2010}), \eprint{arXiv:0911.5004}.

\bibitem[{\citenamefont{Agullo et~al.}(2008)\citenamefont{Agullo,
  Navarro-Salas, Olmo, and Parker}}]{Agullo:2008qb}
\bibinfo{author}{\bibfnamefont{I.}~\bibnamefont{Agullo}},
  \bibinfo{author}{\bibfnamefont{J.}~\bibnamefont{Navarro-Salas}},
  \bibinfo{author}{\bibfnamefont{G.~J.} \bibnamefont{Olmo}}, \bibnamefont{and}
  \bibinfo{author}{\bibfnamefont{L.}~\bibnamefont{Parker}},
  \bibinfo{journal}{Phys.Rev.} \textbf{\bibinfo{volume}{D77}},
  \bibinfo{pages}{124032} (\bibinfo{year}{2008}), \eprint{arXiv:0804.0513}.

\bibitem[{\citenamefont{Haag et~al.}(1967)\citenamefont{Haag, Hugenholtz, and
  Winnink}}]{Haag:1967sg}
\bibinfo{author}{\bibfnamefont{R.}~\bibnamefont{Haag}},
  \bibinfo{author}{\bibfnamefont{N.}~\bibnamefont{Hugenholtz}},
  \bibnamefont{and} \bibinfo{author}{\bibfnamefont{M.}~\bibnamefont{Winnink}},
  \bibinfo{journal}{Commun.Math.Phys.} \textbf{\bibinfo{volume}{5}},
  \bibinfo{pages}{215} (\bibinfo{year}{1967}).

\bibitem[{\citenamefont{Hugenholtz}(1971)}]{Hugenholtz:1971yr}
\bibinfo{author}{\bibfnamefont{N.}~\bibnamefont{Hugenholtz}}
  (\bibinfo{year}{1971}).

\bibitem[{\citenamefont{Bratteli and Robinson}(1981)}]{Bratteli:1981operator}
\bibinfo{author}{\bibfnamefont{O.}~\bibnamefont{Bratteli}} \bibnamefont{and}
  \bibinfo{author}{\bibfnamefont{D.~W.} \bibnamefont{Robinson}},
  \emph{\bibinfo{title}{Operator Algebras and Quantum Statistical Mechanics
  II}} (\bibinfo{publisher}{Springer, New York}, \bibinfo{year}{1981}).

\bibitem[{\citenamefont{Fulling and Ruijsenaars}(1987)}]{Fulling1987135}
\bibinfo{author}{\bibfnamefont{S.}~\bibnamefont{Fulling}} \bibnamefont{and}
  \bibinfo{author}{\bibfnamefont{S.}~\bibnamefont{Ruijsenaars}},
  \bibinfo{journal}{Physics Reports} \textbf{\bibinfo{volume}{152}},
  \bibinfo{pages}{135 } (\bibinfo{year}{1987}).

\bibitem[{\citenamefont{Troost and Van~Dam}(1977)}]{Troost:1977dw}
\bibinfo{author}{\bibfnamefont{W.}~\bibnamefont{Troost}} \bibnamefont{and}
  \bibinfo{author}{\bibfnamefont{H.}~\bibnamefont{Van~Dam}},
  \bibinfo{journal}{Phys.Lett.} \textbf{\bibinfo{volume}{B71}},
  \bibinfo{pages}{149} (\bibinfo{year}{1977}).

\bibitem[{\citenamefont{Troost and van Dam}(1979)}]{Troost:1978yk}
\bibinfo{author}{\bibfnamefont{W.}~\bibnamefont{Troost}} \bibnamefont{and}
  \bibinfo{author}{\bibfnamefont{H.}~\bibnamefont{van Dam}},
  \bibinfo{journal}{Nucl.Phys.} \textbf{\bibinfo{volume}{B152}},
  \bibinfo{pages}{442} (\bibinfo{year}{1979}).

\bibitem[{\citenamefont{Ashtekar et~al.}(2003)\citenamefont{Ashtekar,
  Fairhurst, and Willis}}]{Ashtekar:2002sn}
\bibinfo{author}{\bibfnamefont{A.}~\bibnamefont{Ashtekar}},
  \bibinfo{author}{\bibfnamefont{S.}~\bibnamefont{Fairhurst}},
  \bibnamefont{and} \bibinfo{author}{\bibfnamefont{J.~L.}
  \bibnamefont{Willis}}, \bibinfo{journal}{Class.Quant.Grav.}
  \textbf{\bibinfo{volume}{20}}, \bibinfo{pages}{1031} (\bibinfo{year}{2003}),
  \eprint{gr-qc/0207106}.

\bibitem[{\citenamefont{Halvorson}(2004)}]{Halvorson-2004-35}
\bibinfo{author}{\bibfnamefont{H.}~\bibnamefont{Halvorson}},
  \bibinfo{journal}{Studies in history and philosophy of modern physics}
  \textbf{\bibinfo{volume}{35}}, \bibinfo{pages}{45} (\bibinfo{year}{2004}).

\bibitem[{\citenamefont{Ashtekar and Lewandowski}(2004)}]{Ashtekar:2004eh}
\bibinfo{author}{\bibfnamefont{A.}~\bibnamefont{Ashtekar}} \bibnamefont{and}
  \bibinfo{author}{\bibfnamefont{J.}~\bibnamefont{Lewandowski}},
  \bibinfo{journal}{Class.Quant.Grav.} \textbf{\bibinfo{volume}{21}},
  \bibinfo{pages}{R53} (\bibinfo{year}{2004}), \eprint{gr-qc/0404018}.

\bibitem[{\citenamefont{Rovelli}(2004)}]{Rovelli2004quantum}
\bibinfo{author}{\bibfnamefont{C.}~\bibnamefont{Rovelli}},
  \emph{\bibinfo{title}{Quantum Gravity}}, Cambridge Monographs on Mathematical
  Physics (\bibinfo{publisher}{Cambridge University Press},
  \bibinfo{year}{2004}).

\bibitem[{\citenamefont{Thiemann}(2007)}]{Thiemann2007modern}
\bibinfo{author}{\bibfnamefont{T.}~\bibnamefont{Thiemann}},
  \emph{\bibinfo{title}{Modern Canonical Quantum General Relativity}},
  Cambridge Monographs on Mathematical Physics (\bibinfo{publisher}{Cambridge
  University Press}, \bibinfo{year}{2007}).

\bibitem[{\citenamefont{Hossain et~al.}(2010)\citenamefont{Hossain, Husain, and
  Seahra}}]{Hossain:2010eb}
\bibinfo{author}{\bibfnamefont{G.~M.} \bibnamefont{Hossain}},
  \bibinfo{author}{\bibfnamefont{V.}~\bibnamefont{Husain}}, \bibnamefont{and}
  \bibinfo{author}{\bibfnamefont{S.~S.} \bibnamefont{Seahra}},
  \bibinfo{journal}{Phys.Rev.} \textbf{\bibinfo{volume}{D82}},
  \bibinfo{pages}{124032} (\bibinfo{year}{2010}), \eprint{arXiv:1007.5500}.

\bibitem[{\citenamefont{Hossain and Sardar}(2014)}]{Hossain:2014fma}
\bibinfo{author}{\bibfnamefont{G.~M.} \bibnamefont{Hossain}} \bibnamefont{and}
  \bibinfo{author}{\bibfnamefont{G.}~\bibnamefont{Sardar}}
  (\bibinfo{year}{2014}), \eprint{arXiv:1411.1935}.

\bibitem[{\citenamefont{Rindler}(1966)}]{Rindler:1966zz}
\bibinfo{author}{\bibfnamefont{W.}~\bibnamefont{Rindler}},
  \bibinfo{journal}{Am.J.Phys.} \textbf{\bibinfo{volume}{34}},
  \bibinfo{pages}{1174} (\bibinfo{year}{1966}).

\bibitem[{\citenamefont{Abramowitz and Stegun}(1964)}]{Abramowitz1964handbook}
\bibinfo{author}{\bibfnamefont{M.}~\bibnamefont{Abramowitz}} \bibnamefont{and}
  \bibinfo{author}{\bibfnamefont{I.}~\bibnamefont{Stegun}},
  \emph{\bibinfo{title}{Handbook of Mathematical Functions: With Formulas,
  Graphs, and Mathematical Tables}}, Applied mathematics series
  (\bibinfo{publisher}{Dover Publications}, \bibinfo{year}{1964}).

\bibitem[{\citenamefont{Barbero~G. et~al.}(2013)\citenamefont{Barbero~G.,
  Prieto, and Villaseñor}}]{Barbero:2013lia}
\bibinfo{author}{\bibfnamefont{J.~F.} \bibnamefont{Barbero~G.}},
  \bibinfo{author}{\bibfnamefont{J.}~\bibnamefont{Prieto}}, \bibnamefont{and}
  \bibinfo{author}{\bibfnamefont{E.~J.} \bibnamefont{Villaseñor}},
  \bibinfo{journal}{Class.Quant.Grav.} \textbf{\bibinfo{volume}{30}},
  \bibinfo{pages}{165011} (\bibinfo{year}{2013}), \eprint{arXiv:1305.5406}.

\bibitem[{\citenamefont{Rovelli}(2014)}]{Rovelli:2014gva}
\bibinfo{author}{\bibfnamefont{C.}~\bibnamefont{Rovelli}}
  (\bibinfo{year}{2014}), \eprint{arXiv:1412.7827}.

\bibitem[{\citenamefont{Campo and Obadia}(2010)}]{Campo:2010fz}
\bibinfo{author}{\bibfnamefont{D.}~\bibnamefont{Campo}} \bibnamefont{and}
  \bibinfo{author}{\bibfnamefont{N.}~\bibnamefont{Obadia}}
  (\bibinfo{year}{2010}), \eprint{arXiv:1003.0112}.

\bibitem[{\citenamefont{Schutzhold et~al.}(2006)\citenamefont{Schutzhold,
  Schaller, and Habs}}]{Schutzhold:2006gj}
\bibinfo{author}{\bibfnamefont{R.}~\bibnamefont{Schutzhold}},
  \bibinfo{author}{\bibfnamefont{G.}~\bibnamefont{Schaller}}, \bibnamefont{and}
  \bibinfo{author}{\bibfnamefont{D.}~\bibnamefont{Habs}},
  \bibinfo{journal}{Phys.Rev.Lett.} \textbf{\bibinfo{volume}{97}},
  \bibinfo{pages}{121302} (\bibinfo{year}{2006}), \eprint{quant-ph/0604065}.

\bibitem[{\citenamefont{Schutzhold et~al.}(2008)\citenamefont{Schutzhold,
  Schaller, and Habs}}]{Schutzhold:2008zza}
\bibinfo{author}{\bibfnamefont{R.}~\bibnamefont{Schutzhold}},
  \bibinfo{author}{\bibfnamefont{G.}~\bibnamefont{Schaller}}, \bibnamefont{and}
  \bibinfo{author}{\bibfnamefont{D.}~\bibnamefont{Habs}},
  \bibinfo{journal}{Phys.Rev.Lett.} \textbf{\bibinfo{volume}{100}},
  \bibinfo{pages}{091301} (\bibinfo{year}{2008}).

\bibitem[{\citenamefont{Aspachs et~al.}(2010)\citenamefont{Aspachs, Adesso, and
  Fuentes}}]{Aspachs:2010hh}
\bibinfo{author}{\bibfnamefont{M.}~\bibnamefont{Aspachs}},
  \bibinfo{author}{\bibfnamefont{G.}~\bibnamefont{Adesso}}, \bibnamefont{and}
  \bibinfo{author}{\bibfnamefont{I.}~\bibnamefont{Fuentes}},
  \bibinfo{journal}{Phys.Rev.Lett.} \textbf{\bibinfo{volume}{105}},
  \bibinfo{pages}{151301} (\bibinfo{year}{2010}), \eprint{arXiv:1007.0389}.

\end{thebibliography}

\end{document}